\documentclass[preprint]{aastex631}
\usepackage{bm}
\usepackage{comment}
\usepackage{color}
\usepackage{url}
\submitjournal{ApJS}

\shorttitle{Finite Plasma $\beta$ Magnetic Field Extrapolation}
\shortauthors{Yamasaki et al.}

\begin{document}

\title{Finite Plasma $\beta$ Three-dimensional Magnetic Field Extrapolation Based on MHD Relaxation Method}

\correspondingauthor{Daiki Yamasaki}
\email{yamasaki.daiki@jaxa.jp}

\author{Daiki Yamasaki}
\affiliation{Institute of Space and Astronautical Science, Japan Aerospace Exploration Agency, \\
3-1-1, Yoshinodai, Chuo-ku, Sagamihara 252-5210, Japan}

\author{Takahiro Miyoshi}
\affiliation{Graduate School of Science, Hiroshima University, \\
1-3-1, Kagamiyama, Higashihiroshima 739-8526, Japan}

\author{Satoshi Inoue}
\affiliation{Center for Solar-Terrestrial Research, New Jersey Institute of Technology, \\
University Heights, Newark, NJ 07102-1938, USA}



\begin{abstract}

Three-dimensional (3D) magnetic field in the solar atmosphere provides crucial information to understand the explosive phenomenon such as solar flares and coronal mass ejections. Since it is still hard that we determine the 3D magnetic field from direct observation, a nonlinear force-free field (NLFFF) extrapolation is one of the best modeling methods that provides 3D magnetic field. However, the method is based on zero-$\beta$ assumption, $i.e.,$ the model ignores the gas pressure gradient and gravitational force. The magnetic field based on an NLFFF is not well reconstructed in high-$\beta$ region, such as in chromospheric or lower height layer and in weak field region. To overcome this problem, we need to consider the magnetohydrostatic equilibrium. In this study, we developed a finite plasma $\beta$ magnetic field extrapolation method based on magnetohydrodynamic relaxation method. In our method, we consider a force balance of the Lorentz force and the gas pressure. We tested three different schemes and extrapolated 3D magnetic field using an observational photospheric vector magnetic field of one solar active region, which is a quadrupole complex sunspot group and well studied with an NLFFF. The verification of three schemes is performed by comparing the residual force, and we concluded that our methods reduce $\sim4\%$ of residual force of the previous NLFFF. We also examined the plasma $\beta$ profile along the height and found that, in the core of the active region, plasma $\beta$ reaches a local minimum of $\approx0.01$ in the lower corona with $\beta\approx1$ at the photosphere.

\end{abstract}



\section{Introduction} \label{sec:intr}
The solar atmosphere consists of partially ionized plasmas, and we often observe magnetohydrodynamic (MHD) phenomenon.
One of the most explosive events is a solar flare \citep{Priest2002}. 
Solar flares are the rapid energy release phenomena driven by magnetic reconnection \citep{Shibata2011}, and the energy source of solar flares is widely considered as magnetic energy accumulated in solar active regions \citep[ARs;][]{Toriumi2019}.
Magnetic field in center part of ARs is highly deviated from the potential state, which is the lowest energy state in given boundary condition of the normal component of magnetic field, and we sometimes find a bundle of twisted field lines, magnetic flux ropes \citep[MFRs;][]{Xu2012,Hanaoka2017,Gibson2018}.
To understand the energetics of solar flares, we need to obtain the three-dimensional (3D) magnetic field in the solar atmosphere quantitatively.
Despite the recent progress in diagnostics of magnetic field of MFRs by using spectropolarimetric observations \citep{Yamasaki2023}, the obtained magnetic field is still two-dimensional in formation height of observed line.

A nonlinear force-free field (NLFFF) extrapolation is one of the most efficient methods to reconstruct 3D magnetic field in solar corona \citep{Wiegelmann2012,Inoue2016}.
An NLFFF is extrapolated based on observed photospheric vector magnetic field under a force-free approximation, which has good agreement with extreme-ultraviolet (EUV) observations and it is widely applied to solar flare studies mainly focusing on the formation process of MFRs and change of the magnetic configuration before and after flares \citep{SuY2009,Savcheva2009,Inoue2013,Kang2016,Kawabata2017,Muhamad2018,Yamasaki2021,Teraoka2025}. 
In addition, an NLFFF is also useful to reproduce MFR eruptions in data-based MHD simulations \citep{Jiang2013,Inoue2014b,Inoue2018b,HeW2020,Inoue2021,Yamasaki2022b,LiuN2025,Matsumoto2025X}.

However, an NLFFF is based on zero-$\beta$ assumption, $i.e.,$ the gas pressure gradient and gravitational force are not taken into account, as a result, the magnetic field is not well reconstructed in high-$\beta$ region, such as in chromosphere and in weak field region.
Some previous studies found inconsistency between an NLFFF and observed magnetic field in chromosphere.
\citet{YellesChaouche2012} compared two extrapolation results based on an NLFFF and pointed out that one result which is based on photospheric magnetic field as a bottom boundary is different in terms of number of magnetic twist from the other result which is based on chromospheric magnetic field as a bottom boundary.
\citet{Kawabata2020} indicated that the observed magnetic shear angle from the potential field at chromospheric height is considerable comparing to that measured in an NLFFF which is extrapolated with photospheric magnetic field as bottom boundary.
To overcome this problem, we need to built an extrapolation method based on the magnetohydrostatic (MHS) equilibrium \citep{Miyoshi2020, Zhu2022, Wiegelmann2023}.
\citet{Zhu2020} applied a developed extrapolation code based on MHS to an actual solar active region and found that the MHS field lines are better aligned to fibrils compared to the NLFFF lines.

In this study, for the first step of a development of an extrapolation method based on MHS condition, we developed finite plasma $\beta$ 3D coronal magnetic field extrapolation method by including gas pressure.
We tested three different numerical schemes and performed 3D magnetic field extrapolation using an observational photospheric vector magnetic field as a bottom boundary.
The verification is performed by comparing the residual force.
The rest of this paper is structured as follows:
the numerical method we newly developed is described in Section \ref{sec:meth}, results are presented in Section \ref{sec:resu}, and discussions and summary on the extrapolation results are shown in Section \ref{sec:disc}.

\section{Numerical Method} \label{sec:meth}
\subsection{MHS Equation} \label{subsec:mhseq}
First, we describe the MHS equilibrium by following the explanation given in \citet{Miyoshi2020}.
The basic MHS equation is given by
\begin{eqnarray}
  \bm{J}\times\bm{B}-\bm{\nabla}p-\rho g\bm{e_Z} &=& 0, \label{eq10}\\
  \bm{\nabla}\cdot\bm{B} &=& 0, \\
  p &=& \rho k_BT/m,
\end{eqnarray}
where $\bm{J}$, $\bm{B}$, $p$, $\rho$, $T$, $k_B$, $m$, $g$, and $\bm{e_z}$ are the electric current density, magnetic field, gas pressure, density, temperature, Boltzmann constant, mean molecular weight of gas, gravitational acceleration, and the unit vector in $z$ direction, respectively.
By subtracting the background hydrostatic field, $p_0$ and $\rho_0$, from equation (\ref{eq10}), we obtain
\begin{eqnarray}
  \bm{J}\times\bm{B}-\bm{\nabla}\tilde{p}-\tilde{\rho} g\bm{e_Z}=0, \label{eq11}
\end{eqnarray}
where the components deviated from the hydrostatic equilibrium are respectively defined as
\begin{eqnarray}
  \tilde{p} = p-p_0, ~\tilde{\rho} = \rho-\rho_0, \label{eq12}
\end{eqnarray}
which are termed ``pressure deviation'' and ``density deviation'', and we also use the following equation
\begin{eqnarray}
  \frac{\partial p_0}{\partial z}-\rho_0 g=0. \label{eq13}
\end{eqnarray}
Here we assume that the temperature profile in the solar atmosphere depends only on the height $z$ and coincides with the one determined by the background hydrostatic field as
\begin{eqnarray}
  T(z)=\frac{p_0(z)}{\rho_0(z)}=\frac{p}{\rho}. \label{eq14}
\end{eqnarray}
Hence, 
\begin{eqnarray}
  \tilde{\rho}=\frac{\tilde{p}}{T(z)}. \label{eq15}
\end{eqnarray}
By substituting equation (\ref{eq15}) into equation (\ref{eq10}), we obtain
\begin{eqnarray}
  \bm{J}\times\bm{B}-\bm{\nabla}\tilde{p}-\frac{\tilde{\rho}}{H(z)}\bm{e_Z}=0, \label{eq16}
\end{eqnarray}
where the pressure scale height $H(z)$ is defined as following.
\begin{eqnarray}
  H(z)\equiv\frac{T(z)}{g}. \label{eq17}
\end{eqnarray}
In our extrapolation code, we ignore the gravitational force in equation (\ref{eq10}).
This case is justified when the pressure scale height $H(z)$ is larger or comparable to the size of our numerical domain.
In the case the gravitational force is negligible comparing to the Lorentz force and pressure gradient terms.

We note that, in our calculations, the numerical domain has dimensions of $255\times195\times191$ $\mathrm{Mm^{3}}$, or $1.00\times0.77\times0.75$ in nondimensional units.
The region is divided into $352\times270\times264$ grid points, which corresponds to the grid size of $724$ $\mathrm{km/pix}$.
The typical pressure scale height in transition region or corona is $\sim100$ $\mathrm{Mm}$ by assuming $T$ of order of $10^{6}$ K, this spatial size is large enough to the grid size and comparable to the numerical domain in this study.
In the photosphere and chromosphere, where the temperature is of order $10^{4}$ K, the pressure scale height is $\sim1$ $\mathrm{Mm}$.
Therefore, a realistic modeling of the magnetic field and plasma from the photosphere through the chromosphere into the corona requires inclusion of the gravity term.
In the present paper, however, we focus on optimizing the numerical scheme for the case in which the gravity term is neglected.

\subsection{Basic Equations} \label{subsec:baeq}
As we introduced in Section \ref{subsec:mhseq}, in our extrapolation code, we consider the force balance between the Lorentz force and the gas pressure gradient.
We name this as ``finite plasma $\beta$'' equation.
To obtain the 3D magnetic field, we solve the following finite plasma $\beta$ MHD equations, 
\begin{eqnarray}
  \rho_B &=& |\bm{B}|, \label{eq1}\\
  \frac{\partial \bm{v}}{\partial t} &=& -(\bm{v}\cdot{\bm{\nabla}})\bm{v}+\frac{1}{\rho_B}\bm{J}\times\bm{B}+\nu\bm{\nabla}^{2}\bm{v}-\frac{1}{\rho_B}\bm{\nabla}\tilde{p}, \label{eq2}\\
  \frac{\partial \bm{B}}{\partial t} &=& \bm{\nabla}\times(\bm{v}\times\bm{B}-\eta\bm{J})-\bm{\nabla}\phi, \label{eq3}\\
  \bm{J} &=& \bm{\nabla}\times\bm{B}, \label{eq4}\\
  \frac{\partial \phi}{\partial t} &=& -c_{\mathrm{h}}^{2}\bm{\nabla}\cdot\bm{B}-\frac{c_{\mathrm{h}}^{2}}{c_{\mathrm{p}}^{2}}\phi, \label{eq5}\\
  \frac{\partial \tilde{p}}{\partial t} &=& -\rho_B a^{2}\bm{\nabla}\cdot\bm{v}+\kappa\bm{\nabla}^{2}\tilde{p}-\bm{v}\cdot\bm{\nabla}\tilde{p}, \label{eq6}
\end{eqnarray}
where $\rho_B$, $\tilde{p}$, $\bm{B}$, $\bm{v}$, $\bm{J}$, and $\phi$ are plasma pesudo-density, gas pressure deviation, magnetic flux density, velocity, electric current density, and a conventional potential to reduce errors derived from $\bm{\nabla}\cdot\bm{B}$ \citep{Dedner2002}. 
The pseudo-density in equation (\ref{eq1}) is assumed to be proportional to $|\bm{B}|$.
The assumption is aimed to make the Alfv{\'e}n velocity uniform in the numerical domain.
We note that the gas pressure deviation $\tilde{p}$ in our equations follows the equation (\ref{eq12}).
To obtain absolute value of the gas pressure ($p$), we should consider the consistency of the equation of state in the solar atmosphere when we select the background pressure ($p_0$).
In these equations, the length, magnetic field, density, velocity, time, and electric current density are normalized by $L^*=255$ $\mathrm{Mm}$, $B^*=3000$ $\mathrm{G}$, $\rho^{*}=|B^{*}|$, $V_{\mathrm{A}}^{*}\equiv B^*/(\mu_{0}\rho^*)^{1/2}$, where $\mu_{0}$ is the magnetic permeability, $\tau_{\mathrm{A}}^{*}\equiv L^{*}/V_{\mathrm{A}}^{*}$, and $J^{*}=B^{*}/\mu_{0}L^{*}$, respectively.
Here we note that $V_{\mathrm{A}}^{*}$ and $\tau_{\mathrm{A}}^{*}$ correspond to the Alfv{\'e}n velocity and Alfv{\'e}n time, respectively.
We also note that the normalization values of $255$ $\mathrm{Mm}$ for the spacial scale and $3000$ $\mathrm{G}$ for the magnetic field strength depend on the bottom boundary condition.
In equation (\ref{eq2}), $\nu$ is a viscosity fixed at $1.0\times10^{-3}$.
The resistivity in equation (\ref{eq3}) is given as $\eta=\eta_{0}+\eta_{1}|\bm{J}\times\bm{B}||\bm{v}|^{2}/|\bm{B}|^{2}$, where $\eta_{0}=5.0\times10^{-5}$ and $\eta_{1}=1.0\times10^{-3}$ in non-dimensional units.
The coefficients $c_{\mathrm{h}}^{2}$ and $c_{\mathrm{p}}^{2}$ in equation (\ref{eq5}) are numerical coefficients to achieve divergence $B$ error dissipation and propagation, and they are fixed to the constant values $0.04$ and $0.1$, respectively \citep{Dedner2002}.
In equation (\ref{eq6}), $a$ is a pesudo-sound speed, which follows $a=0.1V_{\mathrm{A}}^{*}$.
The artificial viscosity coefficient $\kappa$ in equation (\ref{eq6}) is given with $\kappa=|\bm{v}|\delta r$, where $\delta r$ is the grid size of numerical domain.
We note that the artificial viscosity $\kappa$ in our equation is designed to mimic the upwind-type scheme.

\subsection{Numerical Schemes} \label{subsec:nusc}
\begin{figure*}[htb]
  \begin{center}
    \includegraphics[bb= 0 0 634 360, width=120mm]{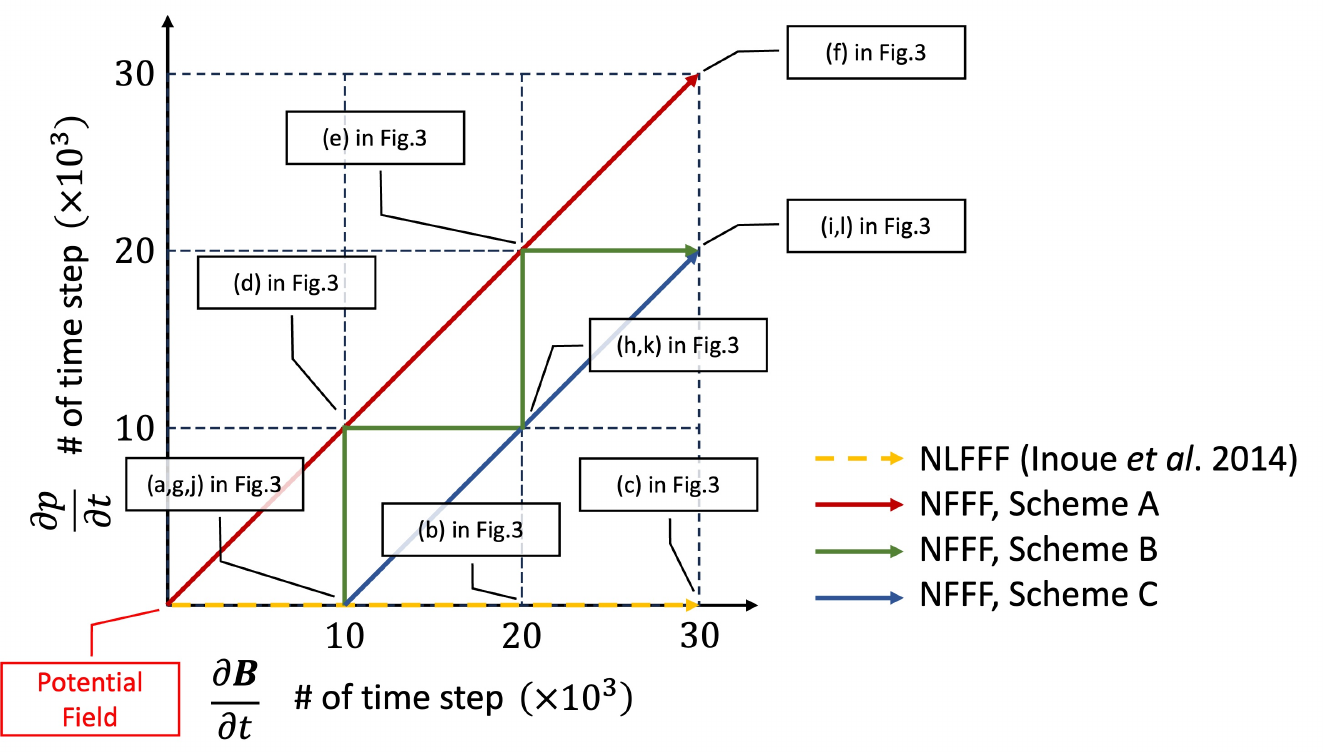}
    \caption{Schematics of three schemes. Horizontal and vertical axes correspond to the numbers of time step of magnetic field evolution and gas pressure evolution, respectively. Red, green, and blue solid arrows show the calculation path of Scheme A, B, and C, respectively. Yellow dashed arrow shows the path of an NLFFF extrapolation.}
    \label{fig1}
  \end{center}
\end{figure*}

We perform experiments on finite plasma $\beta$ 3D magnetic field extrapolation with three different schemes: Schemes A, B, and C.
As shown in red arrow in Figure \ref{fig1}, in Scheme A, we solve all the equations (\ref{eq1}-\ref{eq6}) simultaneously for 30000 time steps.
In Scheme B, as shown in green arrow in Figure \ref{fig1}, we first solve the equations (\ref{eq1}-\ref{eq5}) simultaneously for 10000 time steps, $i.e.,$ we do not solve the evolution of gas pressure during this period.
We, then, solve the equations (\ref{eq1}-\ref{eq6}) except equation (\ref{eq3}) for another 10000 time steps, $i.e.,$ we do not solve the evolution of magnetic field.
We, further, solve the equations (\ref{eq1}-\ref{eq5}) for next 10000 time steps, and this process is repeated.
In Scheme C, as shown in blue arrow in Figure \ref{fig1}, we first solve the equations (\ref{eq1}-\ref{eq5}) simultaneously for 10000 time steps.
We, then, solve all the equations (\ref{eq1}-\ref{eq6}) for next 20000 time steps.
For all the schemes, the second-order central finite difference method and the fourth-order Runge-Kutta-Gill method \citep{Gill1951} are adopted.
We note that the dashed yellow arrow shown in Figure \ref{fig1} represents a calculation path of an NLFFF extrapolation \citep{Inoue2014,Inoue2016}, which we use as a reference.

For all the schemes, we first calculate the potential field \citep{sakurai1982} which is used as the initial condition of the magnetic field of MHD relaxation.
Regarding the initial condition of the gas pressure deviation, we set as the gas pressure deviation follows $\tilde{p}=-\bm{B}^2/2$.
As for the boundary conditions, three components of the magnetic field are fixed at each boundary, while the velocity is fixed to zero and the von Neumann condition $\partial/\partial n=0$ is imposed on $\phi$ and $p$ during the iteration. 
Here we note that we fixed the bottom boundary according to
\begin{eqnarray}
  \bm{B_{\mathrm{bc}}}=\gamma\bm{B_{\mathrm{obs}}}+(1-\gamma)\bm{B_{\mathrm{pot}}}, \label{eq7}
\end{eqnarray}
where $\bm{B_{\mathrm{bc}}}$ is the transversal component determined by a linear combination of the observed magnetic field ($\bm{B}_{\mathrm{obs}}$) and the potential magnetic field ($\bm{B_{\mathrm{pot}}}$).
$\gamma$ is a coefficient in the range of 0 to 1.
The value of the parameter $\gamma$ is increased to $\gamma=\gamma+d\gamma$ if $R=\int|\bm{J}\times\bm{B}|^{2}dV$, which is integrated over the computational domain, becomes smaller than a critical value which is denoted by $R_{\mathrm{min}}$ during the iteration.
In this paper, we set $R_{\mathrm{min}}$ and $d\gamma$ the values of $5.0\times10^{-3}$ and $0.02$, respectively.
When $\gamma$ reaches to 1, $\bm{B_{\mathrm{bc}}}$ is completely consistent with the observed data.
Furthermore, we controll the velocity as follows.
If the value of $v^{*}(=|\bm{v}|/|\bm{v_{A}}|)$ is larger than $v_{\mathrm{max}}$ (here we set to $0.04$), then we modify the velocity from $\bm{v}$ to $(v_{\mathrm{max}}/v^{*})\bm{v}$.
We adopted these processes because they would help avoid a sudden jump from the boundary into the domain during the iterations.

\subsection{Test Data for Verification}
For a test and a verification of our schemes, we used actual solar observation data of the solar active region (AR) NOAA 12887 on 2021 October 28 14:00 UT as a bottom boundary.
We selected the data for the test and verification not only because the AR produced the GOES class X1.0 flare and coronal mass ejection on the same day, but also because we have already well analyzed the AR by using an NLFFF extrapolation \citep{Yamasaki2022b} and found that the MFR was formed at the center part of the AR. 

The photospheric vector magnetic field are taken by the Helioseismic and Magnetic Imager \citep[HMI; ][]{Scherrer2012} onboard the $Solar~ Dynamics~ Observatory$ \citep[$SDO$; ][]{Pesnell2012}.
The vector magnetograms which we use have been released as the Spaceweather HMI Active Region Patch \citep[SHARP; ][]{Bobra2014} data series (hmi.sharp\_cea\_720s series).
Details of the vector magnetic field data reduction and other related information about HMI data products can be found in \citet{Hoeksema2014} and \citet{Bobra2014}.
To compare the extrapolation results of finite plasma $\beta$ magnetic field to a well-studied NLFFF with the same boundary condition, we performed the preprocessing to the vector magnetic field according to \citet{Wiegelmann2006}.

\begin{figure*}[htb]
  \begin{center}
    \includegraphics[bb= 0 0 524 644, width=150mm]{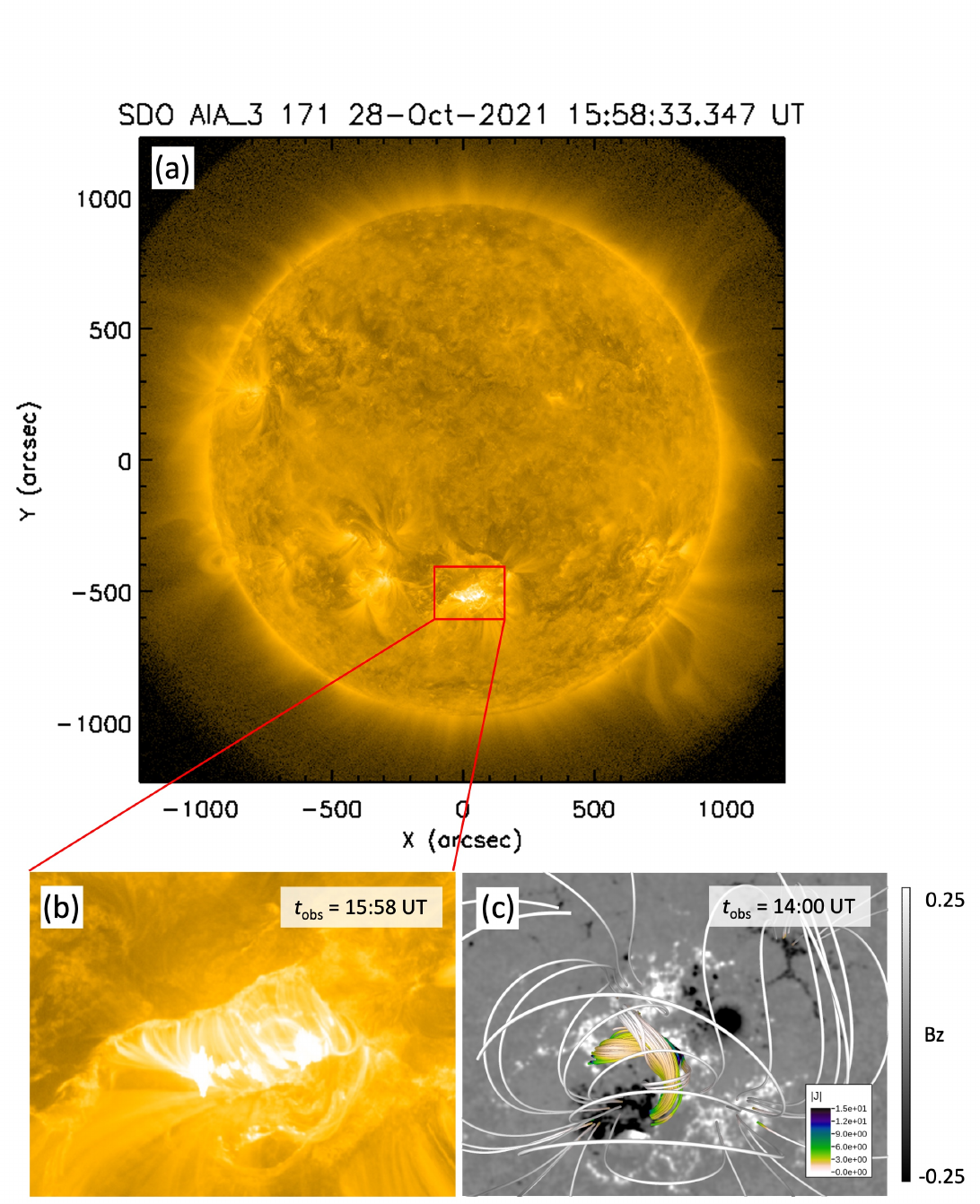}
    \caption{Overview of the target AR which we used as test data for magnetic field extrapolation. (a) Full-disk EUV 171~\AA ~ image taken by $SDO$/AIA on 2021 October 28 15:58 UT. Red box indicates the region of interest (ROI). (b) EUV 171~\AA ~ image of ROI. (c) Magnetic field of ROI. Colored field lines show the NLFFF extrapolation results. The color corresponds to the electric current density. Background grayscale show the normal component of the photospheric magnetic field.}
    \label{fig2}
  \end{center}
\end{figure*}

\section{Results and Verification} \label{sec:resu}
\subsection{Evolution of 3D Magnetic Field} \label{sec:3.1}
\begin{figure*}[htb]
  \begin{center}
    \includegraphics[bb= 0 0 644 535, width=150mm]{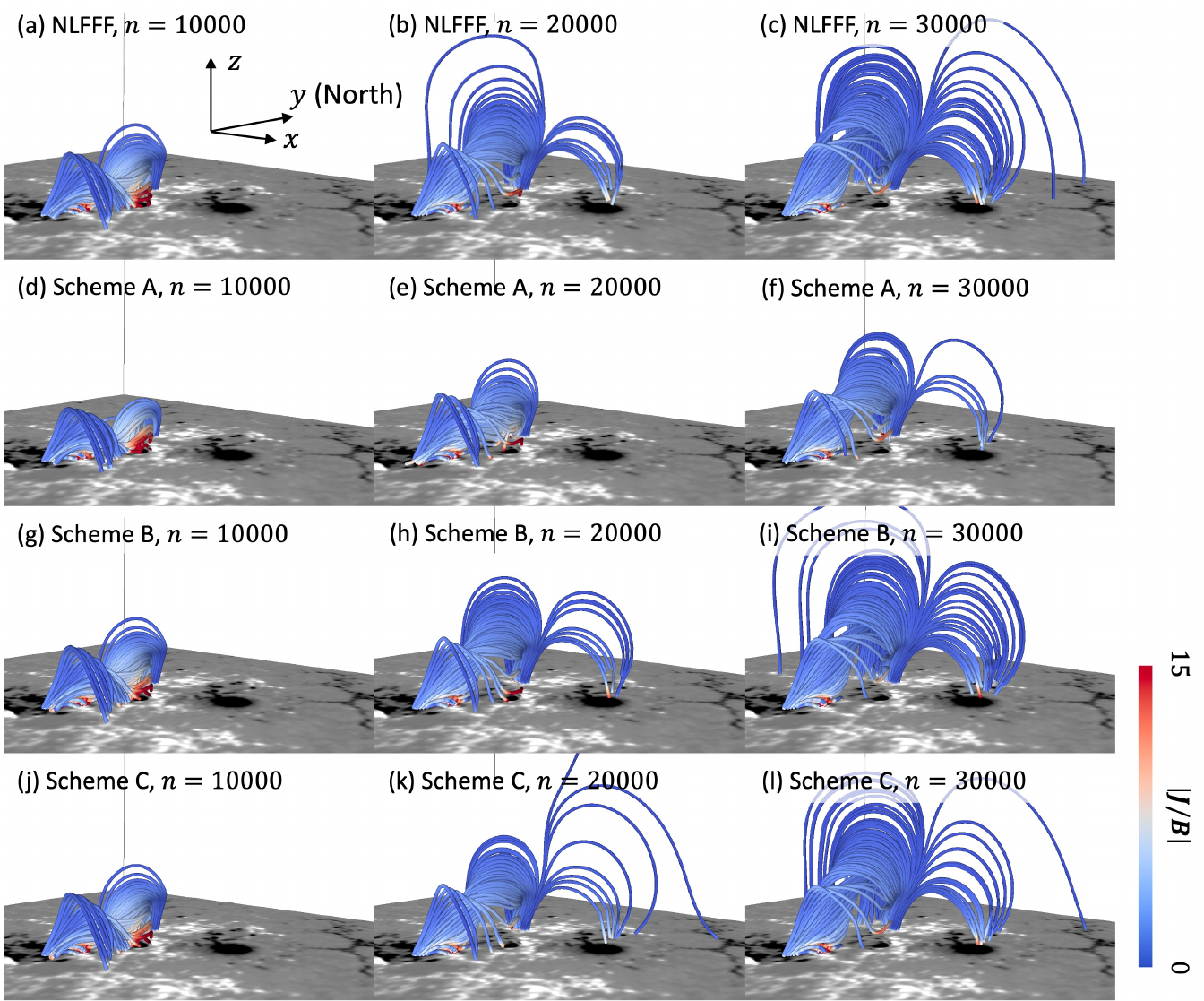}
    \caption{Temporal evolution of the extrapolated magnetic field structures. (a,b,c) time step of 10000, 20000, and 30000 of the NLFFF as a reference. Color of field lines shows the electric current density normalized by the magnetic field strength. The grayscale at the bottom surface ($z=0$) represents the noromal component of the photospheric magnetic field, the color scale is the same as that of panel (c) in Figure \ref{fig2}. (d,e,f) same as (a,b,c) but for Scheme A, respectively. (g,h,i) same as (a,b,c) but for Scheme B, respectively. (j,k,l) same as (a,b,c) but for Scheme C, respectively. (An animation of this figure is available.)}
    \label{fig3}
  \end{center}
\end{figure*}

In Figure \ref{fig3}, we show the extrapolation results for Schemes A, B, and C.
In this study, we used the NLFFF as a reference to verify our schemes of finite plasma $\beta$ extrapolation.
We note that the time step numbers shown in the Figure \ref{fig3} are corresponding to the time step of magnetic field evolution.
In panels (a,b,c), we display the temporal evolution of the magnetic field structure from the potential field to the non-potential field in the case of the NLFFF. 
In panels (d,e,f), (g,h,i), and (j,k,l), we show the results in the case of Schemes A, B, and C, respectively.

Here we note that, as shown in Figure \ref{fig1}, in the case of Schemes B and C, time step number of 10000 is exactly the same as the NLFFF.
Thus, the magnetic field configurations shown in panels (a), (g), and (j) are the same.
We also note that the $\gamma$ value presented in equation \ref{eq7} reached to $1$ before the time step reached 10000 for all the schemes, $i.e.,$ the bottom boundary no longer changes after the time step of 10000.

In panels (f,i,l) which correspond to the extrapolation results with time step of 30000, we find that the highly twisted magnetic field lines are reconstructed in all the schemes as in the NLFFF (see panel (c)).
In the case of Scheme A, the height of field lines are lower than the other cases.

To quantitatively understand the difference of magnetic field properties among the magnetic field obtained from Schemes A, B, C and the NLFFF during the extrapolation, we calculated the temporal evolution of the magnetic free energy ($E_{\mathrm{free}}$) and magnetic twist flux ($\tau$) as defined as follows.
\begin{eqnarray}
E_{\mathrm{free}}&=&\frac{1}{8\pi}\int_{V}(\bm{B}^{2}-\bm{B}_{\mathrm{pot}}^{2})dV, \\
\tau&=&\int_{|T_{\mathrm{w}}|>0.5}|B_{z}|\cdot T_{\mathrm{w}} dS,
\end{eqnarray}
where the magnetic twist is defined as $T_{\mathrm{w}}=(\mu_{0}/4\pi)\int_{L}J_{\parallel}/|\bm{B}|dl$ \citep[cf.][]{Berger2006}. 
We note that the magnetic free energy is integrated over the whole numerical domain, and the magnetic twist flux is integrated around the main polarity inversion line, in which the highly twisted magnetic field lines are found in the NLFFF (see the red box of Figure \ref{fig2} (c)).

In Table \ref{tab1}, we show the results of the evolution of the magnetic free energy and magnetic twist flux.
In all the schemes, both the magnetic free energy and the magnetic twist flux increase by following the iteration number of the magnetic field evolution.
The highest and lowest values of magnetic free energy and the magnetic twist flux is recorded in the case of the NLFFF and Scheme A, respectively.
The differences between the NLFFF and Scheme A are about 10\% in magnetic free energy and about 5\% in magnetic twist flux.
Note that the reason why the physical values are picked up by 30000 iterations is that the magnetic field structure do not change dramatically after that (see the animation of Figure \ref{fig3}).

\begin{table}[htb]
  \begin{center}
  \caption{Evolution of the magnetic free energy and twist flux}
  \label{tab1}
  \begin{tabular}{llrrr}
    \hline
                           &          & \multicolumn{3}{c}{Number of time step} \\
    Item                   & Scheme   & 10000  & 20000  & 30000 \\
    \hline
    Magnetic free energy ($E_{\mathrm{free}}$)           & NLFFF    & $7.2$  & $8.7$  & $9.5$\\
    ($\times10^{33}$ erg)   & Scheme A & $6.4$  & $7.9$  & $8.7$\\
                           & Scheme B & $7.2$  & $8.6$  & $9.3$\\
                           & Scheme C & $7.2$  & $8.5$  & $9.1$\\
    \hline
    Twist flux ($\tau$)             & NLFFF    & $1.6$  & $2.1$  & $2.3$\\
    ($\times10^{6}$ Gauss)  & Scheme A & $1.3$  & $1.8$  & $2.1$\\
                           & Scheme B & $1.6$  & $2.1$  & $2.3$\\
                           & Scheme C & $1.6$  & $2.1$  & $2.2$\\
    \hline   
  \end{tabular}
  \end{center}
\end{table}

\subsection{Evaluation of the Residual Force} \label{sec:3.2}
\begin{figure*}[htb]
  \begin{center}
    \includegraphics[bb= 0 0 980 900, width=120mm]{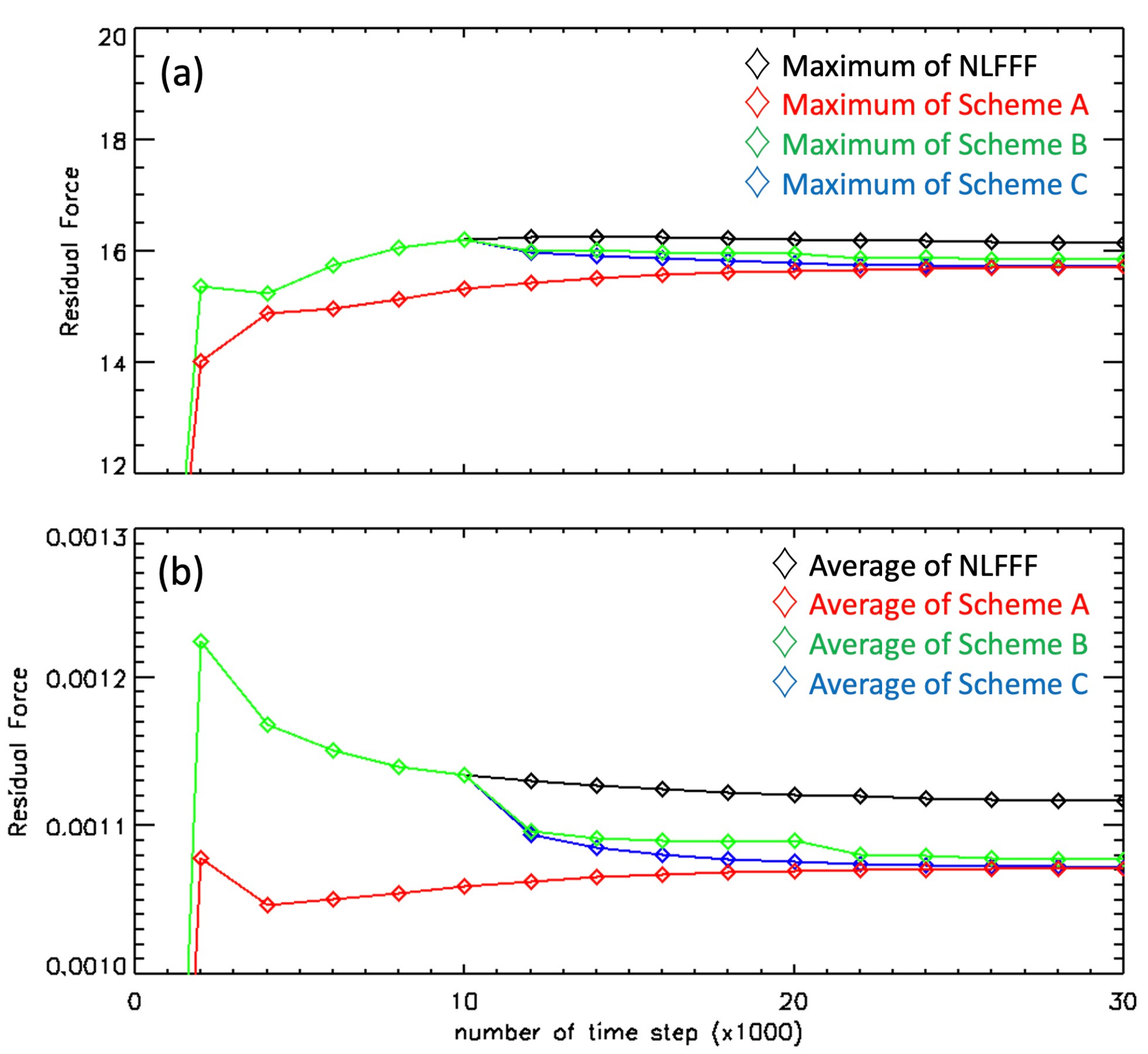}
    \caption{(a) Temporal evolution of the maximum value of residual force in calculation domain. Black, red, green, and blue plots correspond to the NLFFF and the magnetic field obtained from Schemes A, B, and C, respectively. (b) Temporal evolution of an averaged value of each residual force in calculation domain. Color of the plots is the same as that of panel (a).}
    \label{fig5}
  \end{center}
\end{figure*}

In Figure \ref{fig5}, we show the evolution of the residual forces ($F_{\mathrm{res}}$) for the NLFFF and the magnetic field obtained from Schemes A, B, and C.
The residual forces are calculated as following;
\begin{eqnarray}
F_{\mathrm{res}}^{\mathrm{NLFFF}}&=&\frac{1}{\rho_B}\bm{J}\times\bm{B}, \\
F_{\mathrm{res}}^{i}&=&\frac{1}{\rho_B}(\bm{J}\times\bm{B}-\bm{\nabla}p),
\end{eqnarray}
where $i=$ Schemes A, B, and C.
The maximum values and the averaged values of the residual force are shown in panels (a) and (b), respectively.
For all the Schemes A, B, and C, both the maximum and the averaged values of the residual force are lower than that of the NLFFF.
The lowest residual force is in the case of Scheme A, and the value is $\sim4\%$ lower than that of the NLFFF.

\subsection{Magnetic Pressure, Gas Pressure, \& Plasma \texorpdfstring{$\beta$}{beta} Distribution}\label{sec:3.3}
\begin{figure*}[htb]
  \begin{center}
    \includegraphics[bb=0 0 683 354, width=150mm]{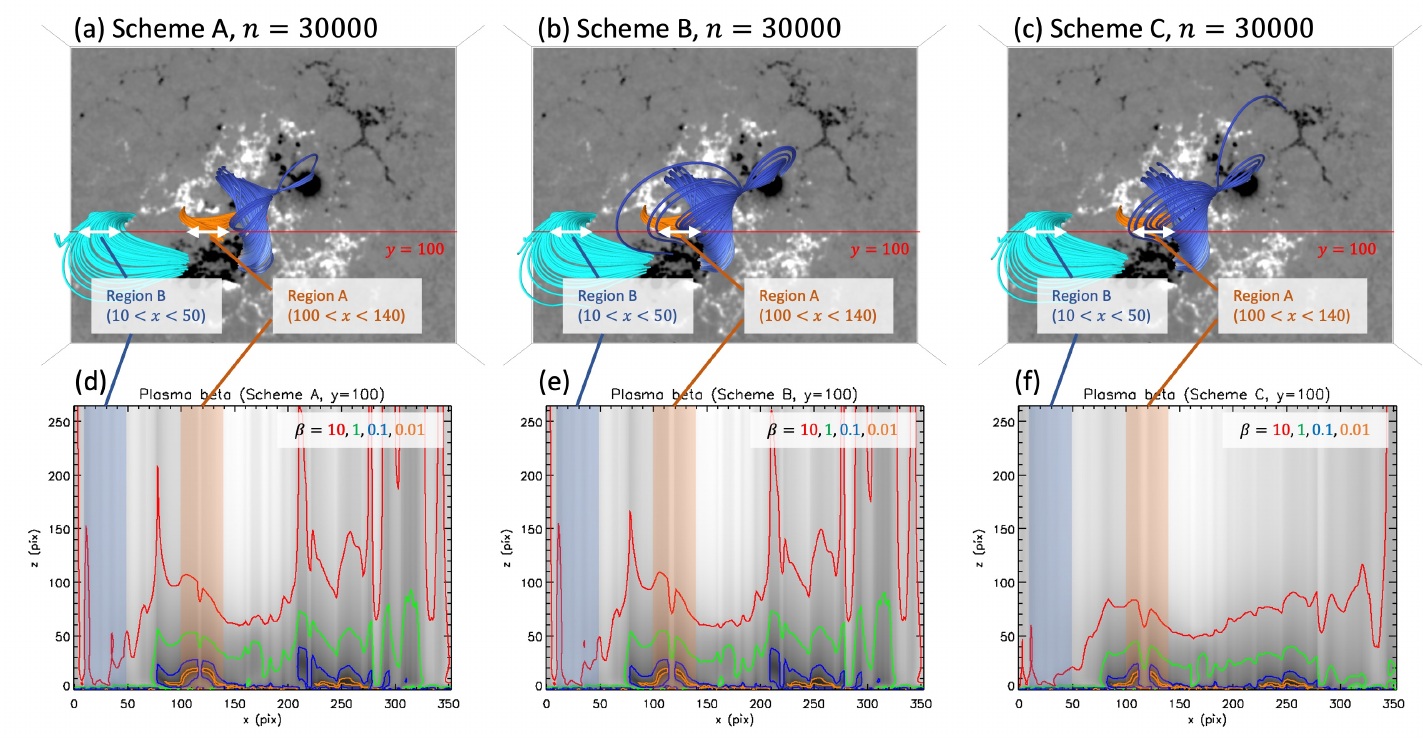}
    \caption{(a) Top view of the magnetic field structure in final-state of Scheme A. Blue field lines are the same as that displayed in Figure \ref{fig3}. Orange and cyan field lines correspond to those around the regions A ($100<x<140, y=100$) and B ($10<x<50, y=100$). (b) Same as (a) but for Scheme B. (c) Same as (a) but for Scheme C. (d) 2-dimensional plasma $\beta$ distribution in $x-z$ cross section ($y=100$) for Scheme A. Red, green, blue, and orange lines display the contours of plasma $\beta$ of 10, 1, 0.1, 0.01, respectively. (e) Same as (d) but for Scheme B. (f) Same as (d) but for Scheme C.}
    \label{fig6}
  \end{center}
\end{figure*}

To confirm the plasma $\beta$ profile along the height direction, we calculated the plasma $\beta$ ($=p_{\mathrm{gas}}/p_{\mathrm{mag}}$), where $p_{\mathrm{gas}}$ ($=\tilde{p}+p_0$) is gas pressure, and $p_{\mathrm{mag}}$ ($=\bm{B}^2/(8\pi)$) is magnetic pressure.
In our analysis, we assumed that the minimum value of the absolute gas pressure, $p_{\mathrm{gas}}$ should be positive in the whole numerical domain, $i.e.,$ we determined the background gas pressure as $p_0=-1\times \mathrm{min}(\tilde{p})+p_{\mathrm{min}}$, where $p_{\mathrm{min}}$ is a small enough constant value that defines the smallest gas pressure in numerical domain, and we adopted $p_{\mathrm{min}}=10^{-6}$ in this study.
To quantitatively compare the plasma $\beta$ profiles between relatively strong and weak magnetic field region, we selected two regions A and B in our dataset.
The region A, $100<x<140, y=100$ in numerical domain, corresponds to the center part of the AR, relatively strong magnetic field region.
The region B, $10<x<50, y=100$ in numerical domain, corresponds to the edge of the AR, relatively weak magnetic field region.
In Figure \ref{fig6} (a), (b), and (c), we show the top view of the magnetic field structure near the regions A and B with orange and cyan lines in the case of Schemes A, B, and C, respectively.
For all the Schemes A, B, and C, we find that the low-lying magnetic field lines in orange around the region A and the potential-like high-lying field lines in cyan around the region B.
In Figure \ref{fig6} (d), (e), and (f), we show the plasma $\beta$ distribution in $x-z$ cross section of $y=100$ in the case of Schemes A, B, and C, respectively.
In the figure, red, green, blue, and orange lines display the contours of plasma $\beta$ of 10, 1, 0.1, and 0.01, respectively.
For all the Schemes A, B, and C, in region A, the plasma $\beta$ value decreases from the photosphere to an altitude of approximately 7–14 Mm (10-20 pix in z-direction in the figure), and then shows an increasing trend with further height.
In contrast, in region B, the plasma $\beta$ value exhibits a monotonic decrease with increasing altitude from the photosphere.

\begin{figure*}[htb]
  \begin{center}
    \includegraphics[bb=0 0 512 384, width=150mm]{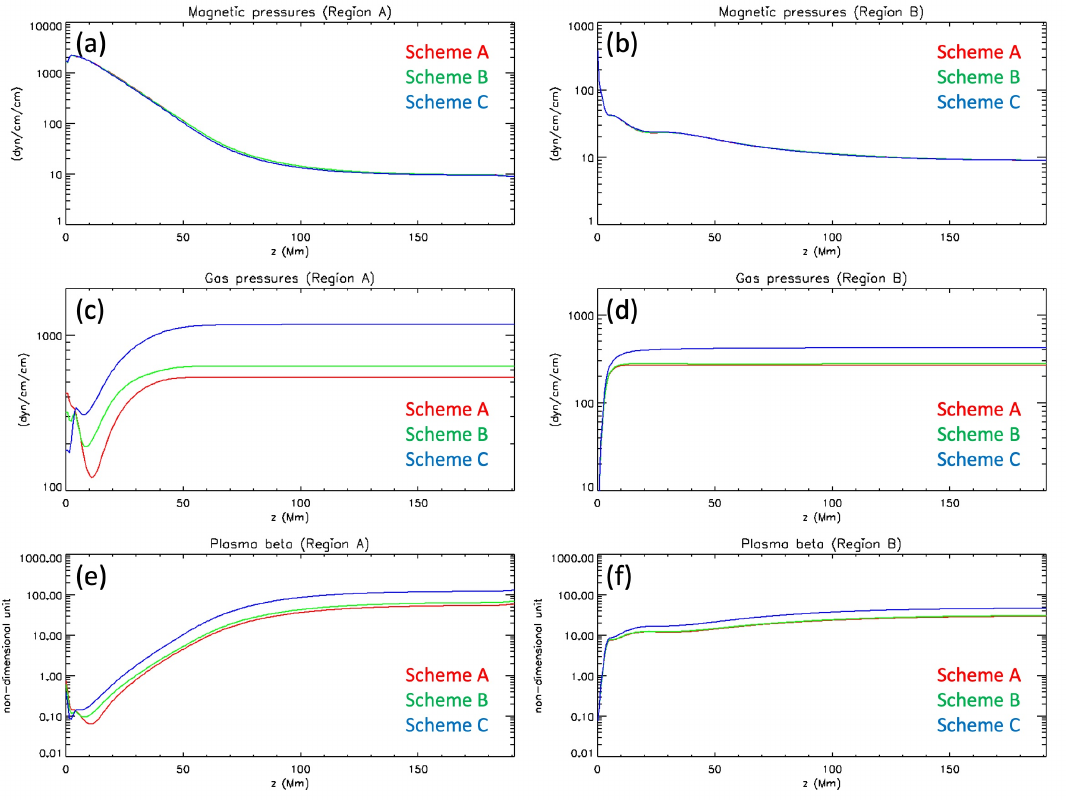}
    \caption{(a) 1-dimensional plot of the magnetic pressure ($p_{\mathrm{mag}}$) in vertical direction of $z$ in averaged over region A. Red, green, and blue plots correspond to the Schemes A, B, and C, respectively. (b) Same as (a) but for region B. (c) Same as (a) but for the gas pressure ($p_{\mathrm{gas}}$) in region A. (d) Same as (c) but for region B. (e) Same as (a) but for the plasma $\beta$ $(=(p_{\mathrm{gas}}/p_{\mathrm{mag}}))$ in region A. (f) Same as (e) but for region B.}
    \label{fig7}
  \end{center}
\end{figure*}

In Figure \ref{fig7}, we show 1-dimensional plots of magnetic pressure ($p_{\mathrm{mag}}$), gas pressure ($p_{\mathrm{gas}}$), and plasma $\beta$ in vertical direction ($z$).
Panels (a), (c), and (e) show the results averaged over region A, while panels (b), (d), and (f) present those for region B.
Red, green, and blue colors correspond to the final-state pressure values of Schemes A, B, and C, respectively.
Regarding the magnetic pressure profiles of Schemes A, B and C in Figure \ref{fig7} (a) and (b), we note that these three plots are so close to each other and they are mostly overlapped.
By comparing the magnetic pressure profiles of regions A and B shown in panels (a) and (b), it is found that in the lower layer (0–50 Mm), region A exhibits significantly higher values than region B, whereas in the upper layer (50–200 Mm), there is no substantial difference between the two regions.
In the gas pressure profiles shown in panels (c) and (d), region A has a local minimum around 10 Mm, while region B shows a steep increase up to 10 Mm and an overall monotonic increase with height.
Comparing the plasma $\beta$ profiles derived from these magnetic and gas pressures, shown in panels (e) and (f), region A, as also noted in Figure \ref{fig6}, has $\beta \approx 1$ near the surface, decreases to a minimum of about 0.01 around 10 Mm, increases back to $\approx 1$ at 30–50 Mm, and then exceeds 10 above approximately 100 Mm.
In contrast, region B shows $\beta \approx 1$ near the photosphere, increases rapidly to about 10 by 10 Mm, and remains $>$10 at higher altitudes.

\begin{figure*}[htb]
  \begin{center}
    \includegraphics[bb= 0 0 697 370, width=150mm]{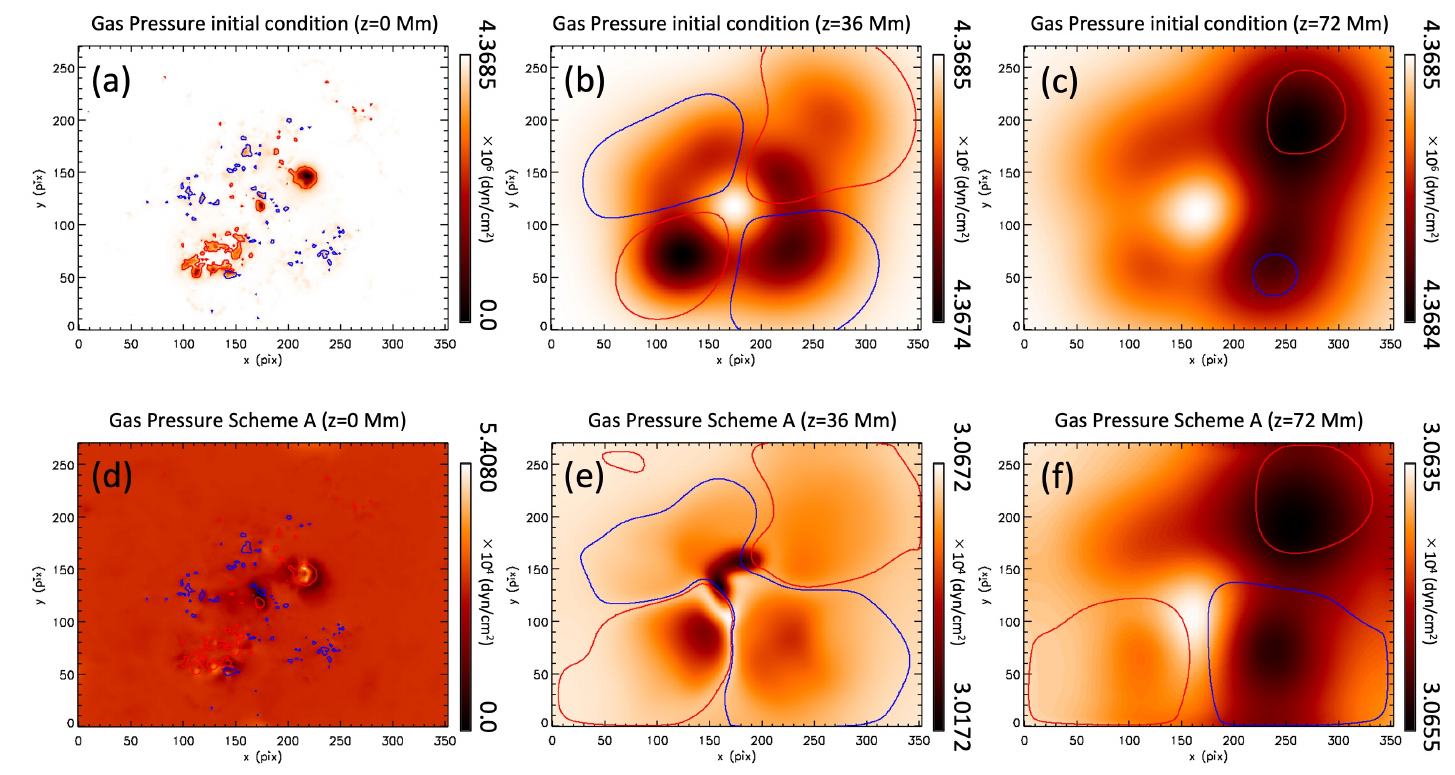}
    \caption{Gas pressure ($p=\tilde{p}+p_0$) distributions of (a) the initial condition with height of $z=0$ Mm, (b) the initial condition with height of $z=36$ Mm, (c) the initial condition owith height of $z=72$ Mm, (d) same as (a) but for the Scheme A final state, (e) same as (b) but for the Scheme A final state, and (f) same as (c) but for the Scheme A final state. For panels (a) and (d), red and blue contours represent the $B_z=\pm750$ $\mathrm{G}$, respectively. For panels (b), (c), (e), and (f), red and blue contours represent the $B_z=\pm75$ $\mathrm{G}$, respectively.}
    \label{fig8}
  \end{center}
\end{figure*}

In Figure \ref{fig8}, we show the spatial distribution of the gas pressure $p$ ($=\tilde{p}+p_0$) in 3 different height of $z=0, 36,$ and $72$ Mm.
We note that the definition of the background gas pressure is the same as what we presented in Figure \ref{fig7} (b).
And since the background gas pressure ($p_0$) was selected artificially under the assumption described above, the absolute value ($p$) is somewhat arbitrary; the physically meaningful quantity is its gradient and the trend of the pressure deviation ($\tilde{p}$).
In panels (a), (b), and (c), the initial condition is displayed, and in panels (d), (e), and (f), the final state of scheme A is shown.
The red and blue contour correspond to the $B_z=\pm750$ $\mathrm{G}$ in the case of panels (a) and (d), and in the case of panel (a), small gas pressure region is found in strong magnetic field region.
This is due to the initial condition of $\tilde{p}=-\bm{B}^2/2$.
However, in the case of panel (d), the gas pressure distribution does not follow the magnetic field strength, and that is far different from the initial condition.
In panels (b), (c), (e), and (f) of Figure \ref{fig8}, the gas pressure is nearly constant at these heights, indicating that the magnetic field is in an almost force-free state.

\section{Discussion and Summary} \label{sec:disc}
As we pointed out in Section \ref{sec:3.1}, in terms of the height of the magnetic field, Scheme A has the lowest height comparing to other cases.
In the case of Scheme A, we solve the evolution of magnetic field and gas pressure simultaneously.
In contrast, in the other cases, only the magnetic field evolution is calculated in the early phase of the extrapolation.   
This result suggests that the gas pressure gradient suppress the Lorentz force which expand the field lines from the lower to higher height.
As shown in Figure \ref{fig1}, the magnetic field structure presented in panels (h) and (k) in Figure \ref{fig3} have the same time steps evolutions in gas pressure and magnetic field.
However, the magnetic field structures of these two are different in the height of magnetic field lines.
Focusing on the dense magnetic field lines in left side, we found that Scheme B (shown in panel (h)) has higher height field lines compared to Scheme C (shown in panel (k)).
In addition, the same trend can be found when we compare the magnetic field structures shown in panels (i) and (l).
These results suggest that once the magnetic field expand by the Lorentz force, it is hard to suppress the height of the magnetic field by the gas pressure gradient in these numerical schemes.
In addition, we suggest the reason why the magnetic free energy and the magnetic twist flux of Scheme A were lower than that of the NLFFF is, in Scheme A, the magnetic field is balanced not only by the Lorentz force, but also the gas pressure and this results in the reduction of the electric current.
When the electric current becomes smaller, the deviation of the magnetic field from the potential field and the magnetic twist becomes also smaller.

As we pointed out in Section \ref{sec:3.2}, the values in the maximum and average of the residual force for all schemes are lower than that of the NLFFF.
Interestingly, in panel (b) of Figure \ref{fig5}, the evolution of averaged value of residual force in Scheme B has sudden decrease between time steps of 10000 and 11000 and between time steps of 20000 and 21000 (see green plots in panel (b)).
These timings are corresponding to the switch of magnetic field evolution to gas pressure evolution or the other way round.
Thus, these results suggest that both the magnetic field and the gas pressure evolved to reduce the residual force in Scheme B.
In addition, Scheme C show much more decrease in residual force comparing to Scheme B (see blue plots in Figure \ref{fig5}).
This result suggests that the simultaneous calculation of gas pressure evolution and magnetic field evolution is better in terms of reducing the residual force.
Furthermore, Scheme A shows the lowest residual force among three cases (see red plots in Figure \ref{fig5}).
This result suggests that it is better to reduce the residual force that we start calculation from the potential field.

As we described in Section \ref{sec:3.3}, the gas pressure distribution differed between the initial condition and the final state of all the schemes, especially for the lower height.
This result supports that the gas pressure deviation evolved by following the equation (\ref{eq6}) and reduced the total residual force.
Regarding the height dependence of the plasma $\beta$ obtained in this study, Region A which has relatively strong magnetic fields shows $\beta \approx 1$ at the photosphere, $\beta \approx 0.01$ at 10 Mm (the lower corona), $\beta \approx 1$ at 50 Mm (the middle corona), and $\beta > 10$ at 100 Mm (the upper corona).
This vertical distribution is generally consistent with the height profile of plasma $\beta$ in active regions presented by \citet{Gary2001}.
In contrast, for Region B, where the magnetic field is relatively weak, plasma $\beta$ is about unity near the photosphere and exceeds 10 above 10 Mm.

When the plasma $\beta$ is decomposed into its components, the magnetic pressure and the gas pressure, their respective height profiles reveal distinct characteristics.
The magnetic pressure in both regions A and B decreases with height, which can be interpreted as a natural consequence of magnetic flux tube expansion, and is therefore physically reasonable.
On the other hand, the gas pressure shows a decreasing trend up to the lower corona in region A but increases at higher altitudes.
These results suggest that, in regions of sufficiently strong magnetic field, the equilibrium determined solely by the Lorentz force and gas pressure gradient leads to a local minimum in gas pressure in the lower corona, which in turn produces a minimum in plasma $\beta$.
In region B, the gas pressure increases monotonically from the photosphere to the upper corona.
This upward increase in gas pressure are considered to arise from the assumption in the present model that the gravitational scale height is infinite.
As a direction for future work, we expect that incorporating gravitational stratification of the system may modify the dynamical equilibrium, particularly in regions with weak magnetic fields.

Though we developed finite plasma $\beta$ 3D magnetic field extrapolation method and verified that the residual force is smaller than the NLFFF calculated with the code presented in \citet{Inoue2014}, the difference of the residual force is still $\sim4\%$.
We suggest that the bottom boundary preprocess \citep{Wiegelmann2006}, which was used in this study was affected this result.
This preprocess method is optimized to minimize the Lorentz force at the bottom boundary under force-free condition.
Therefore, if we consider the optimization of the bottom boundary for finite plasma $\beta$ condition, the residual force may become less than in this study \citep[cf.][]{Zhu2022}. 

Since our equation system does not include the gravitational force, this approach is only suitable for upper chromosphere or much higher region, $i.e.$, for the region where the pressure scale height is large enough comparing to the numerical domain.
For the future work of this study, we will implement gravitational force in our numerical code.
Then, we will be able to take into account the whole solar atmosphere from photosphere to corona.
In addition, we need to validate our extrapolation results.
One possible way is to compare the extrapolated magnetic field structure to EUV observations \citep[cf.][]{Yamasaki2021}.

\begin{acknowledgements}
We thank the anonymous referee for helping us to improve and polish this paper.
$SDO$ is a mission of NASA’s Living With a Star Program.
This work was supported by MEXT/JSPS KAKENHI Grant Number JP23K19078 and JP24K07117.
This work was also supported by the National Science Foundation under grant AGS-2145253 .
Visualization of magnetic field lines are produced by VAPOR (\url{www.vapor.ucar.edu}), a product of the Computational Information Systems Laboratory at the National Center for Atmospheric Research \citep{atmos10090488}.
\end{acknowledgements}

\bibliography{yamasaki2023c}
\bibliographystyle{aasjournal}

\end{document}